\documentclass[10pt,letterpaper]{article}
\usepackage[top=0.85in,left=2.75in,footskip=0.75in,marginparwidth=2in]{geometry}

\usepackage[utf8]{inputenc}

\usepackage{cite}

\usepackage{nameref,hyperref}

\usepackage[right]{lineno}

\usepackage{microtype}
\DisableLigatures[f]{encoding = *, family = * }

\raggedright
\usepackage{changepage}

\usepackage[aboveskip=1pt,labelfont=bf,labelsep=period,singlelinecheck=off]{caption}

\makeatletter
\renewcommand{\@biblabel}[1]{\quad#1.}
\makeatother

\usepackage{lastpage,fancyhdr,graphicx}
\usepackage{epstopdf}
\fancyheadoffset[L]{2.25in}
\fancyfootoffset[L]{2.25in}

\usepackage{color}

\definecolor{Gray}{gray}{.25}

\usepackage{graphicx}

\usepackage{sidecap}

\usepackage{wrapfig}
\usepackage[pscoord]{eso-pic}
\usepackage[fulladjust]{marginnote}
\reversemarginpar

\begin{document}
\vspace*{0.35in}

\begin{flushleft}
{\Large
\textbf\newline{ICU Disparnumerophobia and Triskaidekaphobia: The `Irrational Care Unit’?}
}
\newline

Ari Ercole\textsuperscript{1,*}\\

\bigskip
\bf{1} Ari Ercole, University of Cambridge Division of Anaesthesia, Addenbrooke's Hospital, Hills Road, Cambridge, Cambridgeshire, CB2~0QQ, UK.\\

\bigskip
* ae105@cam.ac.uk

\end{flushleft}

\section*{Abstract}
Whilst evidence-based medicine is the cornerstone of modern practice, it is likely that clinicians are influenced by cultural biases. This work set out to look for evidence of number preference in invasive mechanical ventilatory therapy as a concrete example of subconscious treatment bias.
A retrospective observational intensive care electronic medical record database search and analysis was carried out in adult general, specialist neurosciences and paediatric intensive care units within a tertiary referral hospital. All admitted, invasively mechanically ventilated patients between October 2014 and August 2015 were included. Set positive end-expiratory pressure (PEEP), respiratory rate (RR) and inspiratory pressure (Pinsp) settings were extracted. Statistical analysis using conventional testing and a novel Monte Carlo method were used to look for evidence of two culturally prevalent superstitions: Odd/even preference and aversion to the number 13. Patients spent significantly longer with odd choices for PEEP ($OR=0.16$, $p<2\times10^{-16}$), RR ($OR=0.31$, $p<2\times10^{-16}$) and Pinsp (OR=0.48, $p=2.9\times10^{-7}$). An aversion to the number 13 was detected for choices of RR ($p=0.00024$) and Pinsp ($p=3.9\times10^{-5}$). However a PEEP of 13 was more prevalent than expected by chance ($p=0.00028$).
These findings suggest superstitious preferences in intensive care therapy do exist and practitioners should be alert to guard against other, less obvious but perhaps more clinically significant decision-making biases. The methodology described may be useful for detecting statistically significant number preferences in other domains.\\

\textbf{keywords:} Intensive Care, Data Science, Evidence-Based Practice\\


\section*{Introduction}
Intensive care practice is predicated on clear scientific underpinnings and an evidence-based approach. Furthermore it is a highly quantitative specialty involving physiological measurement and formulation of clearly defined and titratable treatment strategies for the support and optimisation of organ function. Whilst outcomes of course are always to some degree uncertain, the use of extensive investigation and monitoring means that the patients are highly characterised and it would be reasonable to assume that there would be little scope of irrational or superstitious practice.

It is known, however, that clinicians do employ intuitive strategies when making decisions in the intensive care unit \cite{lighthall2015}. Furthermore, superstitious / irrational beliefs are common in society. For example the idea that the number 13 is somehow unlucky is taken sufficiently seriously that many hospitals avoid using that number, for example when numbering operating theatres or beds \cite{grier2018}. Various authors have studied number preferences. By way of example, no evidence for inherently `unlucky dates' has been found in emergency service utilisation \cite{lo2012}, angiography \cite{protty2016} outcome or post-tonsillectomy haemorrhage \cite{kumar2004} despite perceptions to the contrary.

However, such deep-seated but belief systems may have psychologically mediated effects on social behaviour and may have profound implications on health. For example the risks of hospitalisation \cite{scanlon1993} or even death \cite{radun2004, Nayha2002} from transport accidents are increased on Friday 13th against matched control days. The mechanisms may be complex and subtle. Whilst traffic volumes \cite{scanlon1993} may be reduced, analysis of shopping activity did not show significantly differences yet nevertheless hospitalisation rates were increased suggesting that superstitious behaviour is covert and may manifest in a task-specific way at a population level.

At the same time, the use of intuitive strategies may clear to the potential for irrational beliefs to subconsciously influence clinical decision making. The intensive care management of ventilation has clearly defined and well-evidenced physiological goals and offers a test case to assess this. It would therefore be expected that ventilator settings should be made on purely rational grounds of achieving appropriate safe tidal volumes and pressures. Choices of ventilator settings should be smoothly distributed, reflecting variability between patients and with time. Demonstration of strong preferences for particular settings over others is evidence for non-EBM behaviour. Since both odd numbers and the number 13 have pervasive superstitious significance in western culture, this study seeks to identify evidence of statistically significant disparnumerophobia and triskaidekaphobia as primary end-points.

\section*{Methods}
Ventilator setting data was extracted as part of a wider service evaluation of ventilatory practice in intensive care at our institution. Hourly data for set respiratory rate (RR) (81,871 patient-hours), positive end-expiratory pressure (PEEP) (188,592 patient-hours) and set inspiratory pressure (Pinsp) (16,590 patient-hours) were obtained by electronic database query of the hospital electronic medical record system in which all ventilator data is automatically stored. The data was fully anonymised: Patient identifiable and date information was never extracted. All invasive, mechanically ventilated patients admitted to general, neuroscience and paediatric intensive care units between October 2014 and June 2015 were included.

\subsection*{Statistics}
To determine odd/even preference the data was divided to determine the total proportion of patient-hours spent with even and odd RR, PEEP and Pinsp. Statistical comparison is straightforward using a 1-sample proportions test with continuity correction.

To determine whether the number 13 differed in popularity, the proportion of patient hours with RR, PEEP and IP settings equal to 13 was obtained from the data. The null hypothesis that the frequency of 13 is given by the distribution of odd settings. This analysis is not possible with standard statistical tests. Instead, the distribution of settings was modelled as a discrete Gaussian distribution with mean and standard deviation estimates obtained from the odd-numbered data. The sampling distribution for the number 13, under the null, could then be obtained from this distribution by 1,000,000 round Monte Carlo simulations and compared to observation using a 2-tailed $Z$-test.

Analysis was performed with the R statistical programming language (The R Foundation for Statistical Computing, Vienna). Statistical significance was taken as $p<0.05$; $p$-values are presented without correction for multiple comparisons.

\section*{Results}
QQ-plots demonstrated that the RR and IP data were acceptably described by a normal distribution. However the PEEP patient-hours distribution (Figure 1) showed a bimodal distribution with a very pronounced peak at PEEP=5cmH$_2$O, whereas the mean PEEP was 6.7cmH$_2$O (s.d. 2.22cmH$_2$O). To correct this non-normality PEEP=5cmH$_2$O data was masked from subsequent analysis leaving 97,123 patient-hours.

For all three parameters, odd numbered settings were statistically significantly less common than even. The odds ratio for an odd PEEP was the smallest ($OR=0.16$; 95\%$CI=0.159-0.164$, $p<2\times10^{-16}$) with a less pronounced effect for RR ($OR=0.31$; 95\%CI=0.306-0.313, $p<2\times10^{-16}$) and Pinsp ($OR=0.48$; 95\%$CI=0.47-0.49$, $p=2.9\times10^{-7}$)

A RR of 13min$^{-1}$ was found to be selected for 1,034 of 25,371 odd patient-hours. For PEEP this was of 307 of 15,660 odd patient-hours (excluding PEEP=5cmH$_2$O) and for Pinsp, the frequency was 13 of 8,626 odd patient-hours. This was compared to expected rates (s.d.) of 1162 (33.3), 244.7 (15.5) and 24.0 (4.899) respectively from Monte Carlo sampling distribution simulation under the null hypothesis that all odd numbered data was drawn from the same discrete Gaussian distribution. Thus we conclude that the choice of 13 was statistically significantly less popular than expected for RR ($p=0.00024$) and Pinsp ($p=3.9\times10^{-5}$) but not for PEEP where 13 was chosen significantly more frequently than expected ($p=0.00028$).

\section*{Discussion}
The results presented here indicate that odd and even number settings are drawn from statistically distinct distributions with odd numbers chosen significantly less frequently.  Similarly evidence is presented for the number 13 being chosen significantly less frequently for RR and Pinsp than would be expected from a discrete Gaussian distribution of odd numbers. Curiously, for PEEP the number 13 was selected more frequently than expected from chance alone although the number 5 was far more popular than any other choice.

\vspace{.5cm} 
\begin{adjustwidth}{-2in}{0in}
\begin{flushright}
\includegraphics[width=163mm]{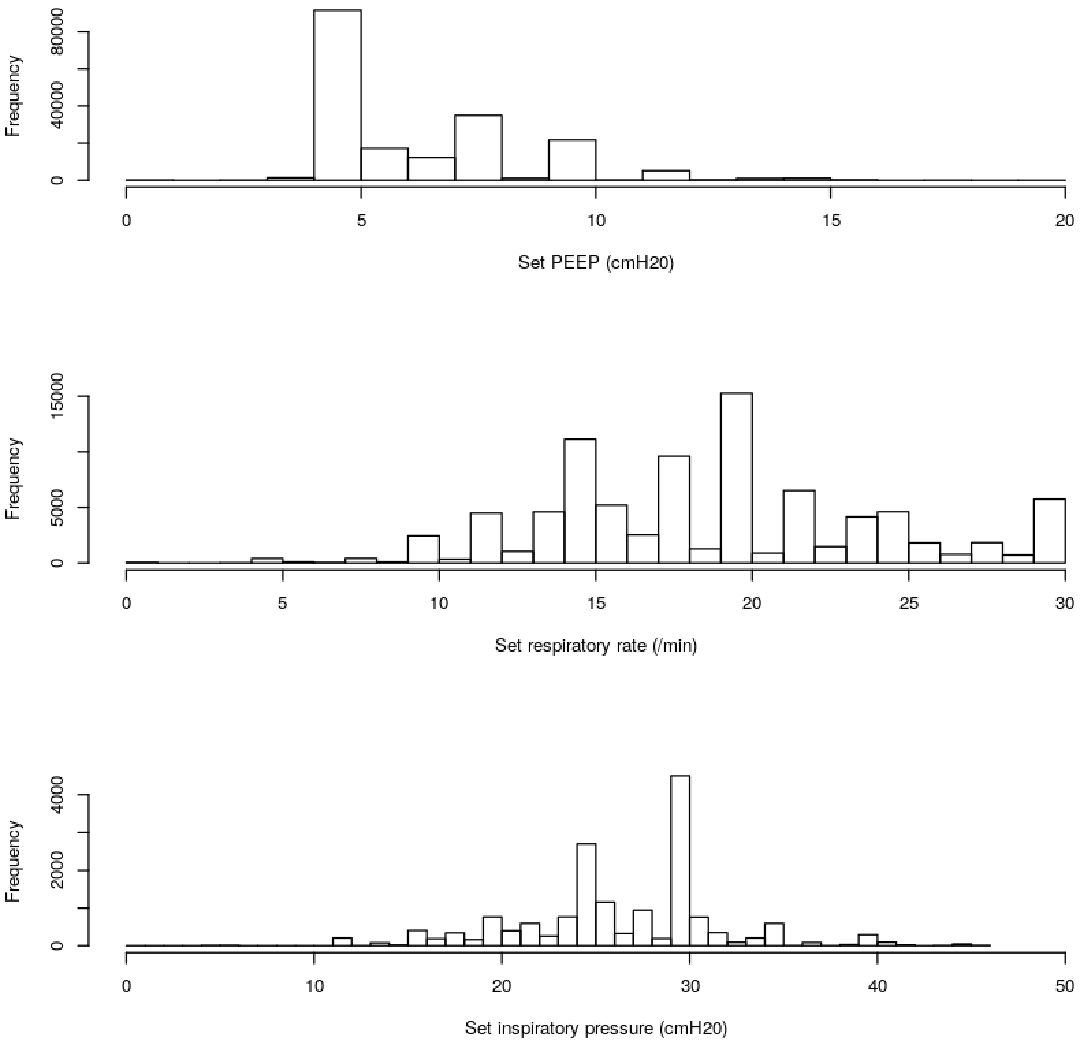}
\label{fig:summary}
\end{flushright}
\justify 
\color{Gray}
\textbf{Figure 1. Summary histograms.} 
Histograms of PEEP, set respiratory rate and set inspiratory pressure.
\end{adjustwidth}

There are two possible interpretations. Either these results indicate irrational choices or there is some biologically plausible mechanism that makes certain numbers more appropriate therapies. In the absence of a credible mechanistic interpretation however, a psychological explanation would seem to be more plausible. Clearly, there is no suggestion that such behaviour is in any way harmful or detrimental to patient care. However it is noteworthy that our number-choices are not entirely based on rational appraisal of physiology.

Of course, PEEP prevents end-expiratory alveolar collapse and therefore shunting. However a PEEP of 5cmH$_2$O is known to be insufficient in preventing alveolar collapse with low tidal ventilation and, should de-recruitment take place, opening pressures are far greater than this pressure \cite{pelosi2001} so this does not justify the choice of the number 5. In our institution, ventilators default to a PEEP of 5cmH$_2$O on start-up, presumably on safety grounds. However the fact that this choice is more prevalent than expected by chance suggests that clinicians choose this value without consideration of the underlying physiology. The `human factors' underlying reluctance to change settings from their default values are important to appreciate if they may bias us to inappropriate ventilatory strategies.

Superstition regarding the unlucky nature of the number 13 in the United Kingdom is widespread. It is more difficult however to understand the finding of `PEEP-triskaidekaphilia’ although this bimodal distribtion may represent a physiologically distribution of maximally haemodynamically or plateau-pressume limited PEEP choices available.

This study includes adult general as well as specialist neuroscience/trauma and paediatric intensive care data. However generalizability is limited by the single tertiary referral hospital site and excludes specialist cardiac intensive care. Additionally, superstitious behaviour is known to be culturally specific and therefore the results cannot be directly extrapolated to other countries or populations.

Implicit in the use of hourly data is that each measurement represents an independent opportunity to adjust and optimise the ventilator. Otherwise, the data is over-sampled, inflating statistical significance. This is justifiable in that nurses typically carry out documentation each hour and thus, presumably, also assess the patient on a similar basis. Indeed assessments may occur more regularly. However, even if this is not the case even if ventilatory decisions were made only once per day (which seems unlikley), this would only reduce our total data by a factor of 24. Given the high statistical significance of our results, even this extreme case would not be expected to change the conclusions although the anonymous data in this work does not allow us to down-sample the data which would need to be done at a patient-level.

This study has focused on mechanical ventilation and thus revealed evidence of numerically irrational practice specifically in intensive care practice. Whether such behaviour is in fact widespread across medicine will require other more universal metrics to be evaluated than instrumentation settings. 

\section*{Conclusions}
Findings consistent with underlying superstitious preferences in the choice of ventilator settings in intensive care are presented. Such preferences are not based in evidence and are irrational. The true prevalence of superstitious preference in medicine is unknown. It seems plausible that clinicians are naturally susceptable to a wide variety of perhaps less easily quantifiable biases and this should be taken into account when designing systems or processes of care.

\section*{Ethical approval}
Data was extracted electronically extracted anonymously as part of a service evaluation into ventilatory strategies and practices in critical care for which local institutional approval was prospectively obtained. Under UK regulations, ethical approval is not required for research conducted using anonymous data obtained during routine clinical practice.

\section*{Acknowledgments}
The author would like to thank Shaun Hyett for help with initial data extraction. The work was not externally funded.

\section*{Conflicts of interest}
None to declare.

\nolinenumbers

\bibliography{library}

\bibliographystyle{vancouver}

\end{document}